\documentclass[apl,twocolumn,a4paper,showpacs,superscriptaddress,floatfix,graphicx,longbibliography]{revtex4-1}

\usepackage{amsmath,amssymb,bm,graphicx}
\usepackage{xcolor}
\usepackage{relsize}
\usepackage[colorlinks=true,urlcolor=blue,linkcolor=blue,citecolor=blue,bookmarks=false]{hyperref}
\usepackage{placeins}

\usepackage[final]{changes}

\begin{document}

\let\oldcite\cite
\renewcommand{\cite}[1]{\mbox{\oldcite{#1}}}

\newcommand{\dMdT}{$\partial M/\partial T$}
\newcommand{\MoSi}{MoSi$_{2}$}
\newcommand{\RT}{$ R_{\mathrm{T}}$}
\newcommand{\dRdT}{$\partial R_{\mathrm{T}}/\partial T$}
\newcommand{\DT}{$ \Delta T $}
\newcommand{\meanT}{$ T_{0}$}
\newcommand{\mstar}{$m^{\ast}$}
\newcommand{\Aone}{$\alpha_{1}$}
\newcommand{\Atwo}{$\alpha_{2}$}
\newcommand{\fDT}{$ \Delta T \cdot f$}
\newcommand{\Rox}{RuO$_{2}$}


\title{Cyclotron mass-selective de~Haas--van~Alphen measurements using temperature modulation}

\author{Michelle Hollricher}
\email{michelle.hollricher@tum.de}
\affiliation{School of Natural Sciences, Technical University of Munich, Garching, Germany}
\affiliation{Zentrum f\"ur QuantumEngineering (ZQE), Technical University of Munich, Garching, Germany}

\author{Andreas Bauer}
\affiliation{School of Natural Sciences, Technical University of Munich, Garching, Germany}
\affiliation{Zentrum f\"ur QuantumEngineering (ZQE), Technical University of Munich, Garching, Germany}

\author{Leo Maximov}
\affiliation{School of Natural Sciences, Technical University of Munich, Garching, Germany}
\affiliation{Zentrum f\"ur QuantumEngineering (ZQE), Technical University of Munich, Garching, Germany}

\author{Louw Feenstra}
\affiliation{School of Natural Sciences, Technical University of Munich, Garching, Germany}
\affiliation{Fakult\"at für Physik, Ludwig-Maximilians-Universit\"at, Munich, Germany}

\author{Christian Pfleiderer}
\affiliation{School of Natural Sciences, Technical University of Munich, Garching, Germany}
\affiliation{Zentrum f\"ur QuantumEngineering (ZQE), Technical University of Munich, Garching, Germany}
\affiliation{Heinz Maier-Leibnitz Zentrum (MLZ), Technical University of Munich, Garching, Germany}
\affiliation{Munich Center for Quantum Science and Technology (MCQST), Technical University of Munich, Munich, Germany}

\author{Marc A. Wilde}
\email{marc.wilde@tum.de}
\affiliation{School of Natural Sciences, Technical University of Munich, Garching, Germany}
\affiliation{Zentrum f\"ur QuantumEngineering (ZQE), Technical University of Munich, Garching, Germany}

\date{\today}

\begin{abstract}
We present a temperature-modulated de~Haas--van~Alphen measurement technique that allows selective addressing of quantum oscillations with different effective masses \mstar\ using a non-monotonic amplitude evolution with temperature and magnetic field, governed by the temperature derivative of the Lifshitz-Kosevich factor. The technique relies on harmonic modulation of the sample temperature and phase-sensitive detection of quantum oscillations in the voltage induced in a pick-up coil. We use a set of frequencies with strong Zeeman-driven harmonic content in the compensated topological semimetal \MoSi\ as a natural linear mass comb ranging from 1\,\mstar\ to 13\,\mstar\ to demonstrate the tunability of the mass-dependent quantum oscillation amplitudes experimentally. \replaced{The technique allows to reliably isolate weak contributions of heavy orbits that are inaccessible in conventional de~Haas--van~Alphen frequency spectra because their frequency peaks overlap with much stronger frequency peaks of lighter orbits.}{Our results establish temperature modulation of the de~Haas--van~Alphen effect as a versatile method capable of isolating weak contributions from heavy orbits that may otherwise be masked by strong oscillations from lighter orbits in the same frequency range.} 
\end{abstract}

\maketitle


The electronic properties of metals are determined by their band structure and Fermi surface. Quantum oscillation measurements, especially the de~Haas--van~Alphen \added{(dHvA) }effect \cite{shoenbergMagneticOscillationsMetals1984,dehaasDependenceSusceptibilityDiamagnetic1930,landauDiamagnetismusMetalle1930}, are among the most powerful tools to probe Fermi surface properties and also provide a precise measure of orbit-specific quasiparticle masses. Experimental challenges arise in materials with complex Fermi surfaces, exhibiting a large number of extremal orbits \replaced{with overlapping frequency peaks on the one hand, and}{ associated} with \added{strongly }different effective masses \mstar\ \added{on the other hand}. Oscillation amplitudes generally decrease with increasing mass, \replaced{making heavy orbits harder to detect than light orbits. Heavy orbits may thus evade detection due to overlap with close-by frequency peaks that are not resolved individually because of the finite magnetic field window or simply due to limitations of the detection system (see supplementary materials).}{so that heavy orbits are easily masked by the much larger oscillations of light orbits in the same frequency range.} However, especially the heavy quasiparticles strongly contribute to the density of states and play a key role in correlated states, including high-temperature superconductivity, heavy fermions, quantum criticality, unconventional magnetism, and correlated topological semimetals.

Prominent topological examples include kagome metals, which host both light Dirac bands and flat bands with correspondingly large mass \cite{linFlatbandsEmergentFerromagnetic2018, kangDiracFermionsFlat2020,boseOriginFlatBands2025,wangQuantumStatesIntertwining2023}. Heavy-fermion materials \cite{andres4fVirtualBoundStateFormationCeAl31975,steglichSuperconductivityPresenceStrong1979,ottUBe13UnconventionalActinide1983,stewartHeavyfermionSystems1984,tailleferDirectObservationHeavy1987} exhibit strong quasiparticle mass enhancements up to several 100\,$m_{\mathrm{e}}$, often accompanied by light Fermi surface pockets \cite{aokiFermiSurfaceProperties2014,liTwodimensionalFermiSurfaces2014,zhangCriticalRoleMagnetic2022}. Chiral B20 materials, known to host multifold fermions and topological nodal planes \cite{huberNetworkTopologicalNodal2022,huberFermiSurfaceMagnetic2025,guoQuasisymmetryprotectedTopologySemimetal2022,klotzElectronicBandStructure2019}, include magnetic representatives such as MnSi \cite{nagaosaTopologicalPropertiesDynamics2013,bauerGenericAspectsSkyrmion2016,wildeSymmetryenforcedTopologicalNodal2021}, where strong coupling to spin fluctuations \cite{lonzarichMagneticOscillationsQuasiparticle1988} leads to large mass enhancements and coexistence of light and heavy masses between $0.1\,m_{\mathrm{e}}-18\,m_{\mathrm{e}}$. Even elemental 3d magnets such as bcc Fe exhibit masses from $0.5\,m_{\mathrm{e}}-10\,m_{\mathrm{e}}$ \cite{lonzarichBandStructureMagnetic1984}. 

In all material classes, \replaced{individually resolving overlapping frequency contributions}{disentangling the signal contributions} of heavy and light quasiparticles is challenging, and measurement approaches that offer selectivity with respect to \mstar\ are thus beneficial.

Topological semimetals are a particularly diverse group known for hosting topologically protected band crossings \cite{armitageWeylDiracSemimetals2018,lvExperimentalPerspectiveThreedimensional2021}. The compensated semimetal \MoSi\ has recently attracted renewed interest due to its non-zero Berry curvature and extremely large magnetoresistance (XMR) \cite{vanruitenbeekHaasvanAlphenEffect1987,matinExtremelyLargeMagnetoresistance2018,pavlosiukGiantMagnetoresistanceFermisurface2022,bhattacharyyaComparisonFullyRelativistic1985,labordeResistivityMagnetoresistanceHighpurity1986,navaAnalysisElectricalResistivity1989,kurganskiiFermiSurfaceElectrical2003}. Its quantum oscillation spectrum is surprisingly rich despite a relatively simple Fermi surface \cite{vanruitenbeekHaasvanAlphenEffect1987,matinExtremelyLargeMagnetoresistance2018,pavlosiukGiantMagnetoresistanceFermisurface2022}, because it exhibits unusually strong harmonic content driven by the interplay of \added{low-curvature }Fermi surface geometry and the Zeeman effect generating spectral contributions with more than 10 times the cyclotron mass of the fundamental orbit.

Our experimental approach uses an oscillating sample temperature \added{$T$} to detect quantum oscillations in the temperature derivative of the magnetization, \dMdT, so that the demodulated signal is proportional to the temperature derivative \dRdT\ of the thermal reduction factor \RT\ in the Lifshitz-Kosevich formalism \cite{lifshitzTheoryMagneticSusceptibility1956,shoenbergMagneticOscillationsMetals1984}. While other quantum oscillation measurement techniques exhibit similar \dRdT\ dependence in principle \replaced{\cite{huangNonlinearShubnikovdeHaas2021,palaciomoralesThermoelectricPowerQuantum2016,mercureQuantumOscillationsMetamagnetic2010,xuCrystalGrowthQuantum2019}}{\cite{palaciomoralesThermoelectricPowerQuantum2016,mercureQuantumOscillationsMetamagnetic2010,xuCrystalGrowthQuantum2019}}, their cyclotron mass-selective properties have not been investigated.

In this letter, we demonstrate the cyclotron mass-selectivity of the temperature-modulated dHvA (TM-dHvA) technique by analyzing the non-monotonic behavior of \dRdT\ with both magnetic field and temperature, and by showcasing the effect in the compensated semimetal \MoSi. The strong harmonic content in its quantum oscillation spectrum forms a natural linear mass comb, where integer multiples in the fundamental frequencies are associated with integer multiples of the fundamental effective mass. \MoSi\ therefore provides an ideal platform for demonstrating the mass-selective capabilities of TM-dHvA.\added{ For a showcase of \emph{overlapping} frequency components with strongly different masses, we refer to the supplementary materials.}


The dHvA effect, as measured in conventional experiments, is a superposition of all oscillatory components $M_i$ associated with the extremal orbits $i$. Each component may be written as
\begin{eqnarray}
\begin{array}{c}
\ M_{i} = \sum_p M_{0,p,i} \,  \, R_{\mathrm{T},p,i}  \mbox{ ,} \\ 
\\
\mathrm{with}\,\,\,\,R_{\mathrm{T},p,i} = \dfrac{x}{\sinh{x}} \mathrm{,}\,\,\,x = \dfrac{2 \pi^{2} p m^{\ast}_{i} k_{\mathrm{B}} T}{\hslash e B} \mbox{ ,}
\end{array}\label{Eq.4}
\end{eqnarray}
where $M_{0,p,i}$ is the $p$th Fourier component of the oscillatory magnetization at zero temperature, \RT $_{p,i}$ is the Lifshitz--Kosevich factor \cite{lifshitzTheoryMagneticSusceptibility1956,shoenbergMagneticOscillationsMetals1984}, \replaced{$m_i^*$}{\mstar} is the cyclotron effective mass corresponding to orbit $i$ and $p$ is the harmonic index.

\begin{figure}[!bt]
\includegraphics{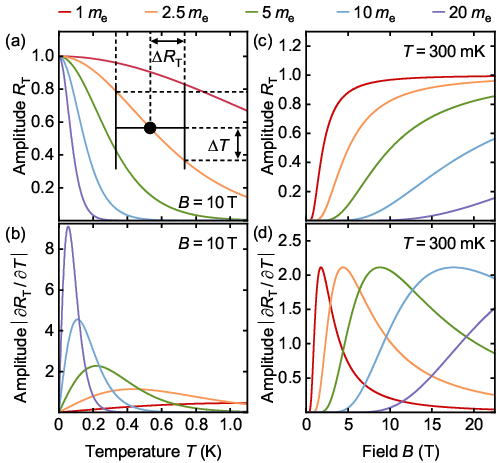}
\caption{Field and temperature dependence of the Lifshitz--Kosevich factor \RT\ and its derivative \dRdT. (a)~\RT\ vs $T$ at 10\,T for selected \mstar. In a conventional dHvA experiment, light \mstar\ tend to dominate the dHvA spectrum at all $T$. Modulation with amplitude $\Delta T$ results in oscillations of \RT\ with amplitude $\Delta$\RT. The demodulated induced voltage $U$ scales with \dRdT. (b)~ \dRdT\ vs $T$. Oscillations with light \mstar\ are strongly suppressed at \emph{low} $T$, where oscillations with heavy \mstar\ dominate and vice versa. (c)~\RT\ vs $B$ at 300\,mK. In a conventional dHvA experiment, light \mstar\ dominate the signal at all $B$. (d)~\dRdT\ vs $B$. Oscillations with light \mstar\ are strongly suppressed at \emph{high} $B$, where oscillations with heavy \mstar\ dominate.\label{Fig:LK}}
\end{figure}

\RT\ is shown in Fig.\,\ref{Fig:LK}(a) for different effective masses ranging from $m^\ast = 1\,m_{\mathrm{e}}$ to $m^\ast = 20\,m_{\mathrm{e}}$  at a constant magnetic field $B=10$\,T. Given that the amplitude of $M_{0,p,i}$ scales inversely with \mstar, it follows that in a conventional dHvA experiment, light masses tend to dominate the spectra at any temperature.


The temperature-modulated dHvA effect has been sketched theoretically in Shoenberg’s seminal book \cite{shoenbergMagneticOscillationsMetals1984} and was applied in \added{Refs.\,}\cite{oderTemperatureoscillationMethodObserving1965,oderNewThermalwaveTechnique1965}. In TM-dHvA measurements, the sample temperature $T$ is modulated at a frequency $f$ so that
\begin{equation}
T = T_{0} + \Delta T \cos{\left( 2 \pi f t \right)} \mbox{ ,}
\label{Eq.2}
\end{equation}
where \meanT\ is the average sample temperature and \DT~is the temperature modulation amplitude. Corresponding changes in the sample magnetization over time induce a voltage in a surrounding pick-up coil via Faraday's law of induction. The voltage $U_{\mathrm{ind}}$ induced in the pick-up coil is proportional to the temperature derivative of the magnetization \dMdT
\begin{equation}
U_{\mathrm{ind}} \sim \dfrac{dM}{dt} \sim \dfrac{\partial M}{\partial T}  \dfrac{dT}{dt}\sim - \dfrac{\partial M}{\partial T} f \Delta T \sin{\left( 2\pi f t \right)}  \mbox{ .}
\label{Eq.3}
\end{equation}

Demodulating  $U_{\mathrm{ind}}$  at frequency $f$  results in a voltage proportional to $\partial M/\partial T$. As sketched in Fig.\,\ref{Fig:LK}(a), modulating the sample temperature with an amplitude $\Delta T$ causes \RT\ to oscillate with an amplitude $\Delta$\RT. The demodulated voltage $U_{i}$ for each extremal orbit $i$ is thus
\begin{equation}
U_{i} \sim f \Delta T \sum_p M_{0,p,i}  \left| \dfrac{\partial R_{\mathrm{T},p,i}}{\partial T} \right| \mbox{ ,}
\label{Eq.5}
\end{equation}
where
\begin{equation}
\added{\left| \frac{\partial R_{\mathrm{T},p,i}}{\partial T} \right| =\frac{x \left(x*\coth{x} -1\right)\text{csch}\,x}{T}}
\end{equation}
\deleted{\dRdT\ }is the temperature derivative of \RT\ shown in Fig.\,\ref{Fig:LK}(b). Further details may be found in Refs.\,\cite{oderNewThermalwaveTechnique1965,shoenbergMagneticOscillationsMetals1984}.


The dependence of the measured signal on \dRdT\ is the reason for the cyclotron mass-selectivity. The \dRdT\ terms corresponding to the same masses as in Fig.\,\ref{Fig:LK}(a) are shown in Fig.\,\ref{Fig:LK}(b). They are non-monotonic and strongly peaked at different temperature values that shift to lower $T$ with increasing \mstar. At sufficiently low temperatures, contributions from light masses vanish, while those from heavy masses are maximized. In contrast to \RT , whose maximum value is one, the maximum value of \dRdT\ has no upper bound, leading to a strong signal enhancement for heavy masses or high harmonics $p$ as compared to a conventional dHvA measurement technique.

Figure\,\ref{Fig:LK}(c) shows \RT\ at $T = 300$\,mK as a function of $B$. The Lifshitz--Kosevich terms increase monotonically from zero to one with the magnetic field. Conventional oscillation amplitude spectra are therefore dominated by light masses at all $B$. The TM-dHvA signal depends on \dRdT $(B)$ as shown in Fig.\,\ref{Fig:LK}(d). Here, heavy masses can be selectively addressed in high magnetic fields, where contributions with light mass are strongly suppressed. Thus, the TM-dHvA technique allows singling out heavy-mass components while simultaneously suppressing light-mass components, which is not possible with quantum oscillation signals scaling with \RT.


\begin{figure}[!tb]
\centering
\includegraphics[width=8.4cm]{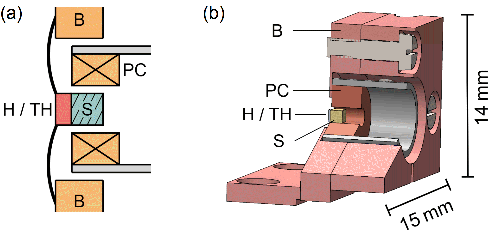}
\caption{TM-dHvA setup. (a) Schematic side view showing the sample~(S) attached to a heater~(H) and thermometer~(TH) platform, thermally linked to the bath~(B). The sample is situated inside the pick-up coil~(PC). (b)~Cutaway view of the setup showing the sample~(S), mounted on a modified bare \Rox~chip that is used as both heater~(H) and thermometer~(TH), situated inside the pick-up coil~(PC).\label{Fig:Setup}}
\end{figure}

The TM-dHvA setup is shown in Fig.\,\ref{Fig:Setup}. It consists of a sample (S), a heater (H) and thermometer (TH), which generate and monitor a sinusoidal oscillation of the sample temperature according to Eq.\,\ref{Eq.2}. The sample is situated inside a pick-up coil (PC) used to inductively detect the resulting oscillation of the sample magnetization. Sample, heater, and thermometer are thermally coupled to the bath (B). In contrast to magnetization measurements that involve oscillating secondary fields, the TM-dHvA setup does not require a secondary coil for signal balancing, parasitic background from the sample environment is negligible, and Eddy current heating is absent.

\deleted{The sample is a 1\,mm cube of single-crystal MoSi$_2$. Starting from stoichiometric amounts of Mo (3N8) and Si (6N), polycrystalline rods were prepared using an arc-melting furnace followed by an inductively heated rod casting furnace \cite{bauerUltrahighVacuumCompatible2016}. Single-crystal growth was carried out in a high-temperature floating-zone furnace at a rate of 10 mm/h under an argon atmosphere of 10 bar \cite{bauerCompositionalStudiesMetals2022}.}

\replaced{A 1\,mm cube of single-crystal MoSi$_2$}{The sample} was attached to the back of a commercially available Lake Shore RX-102A temperature sensor using Apiezon grease. The sensor was modified by cutting a narrow separating trench through the ruthenium oxide layer \cite{michonThermodynamicSignaturesQuantum2019,yangUnveilingDoublepeakStructure2023}. One side is used as the heater, driving the temperature oscillations. The other side is used as the thermometer to monitor both the mean sample temperature \meanT\ and the amplitude of the temperature modulation \DT. The resistance ratio of the heater and thermometer is close to one. The sapphire substrate of the sensor was polished to approximately 200\,$\mu$m to minimize its heat capacity and maximize thermal contact between the heater/thermometer layer and the sample. The electrical connections to the heater and thermometer were made with manganin wires of 30\,$\mu$m diameter.

The pick-up coil~(PC) with approximately 2100 windings is made of enameled copper wire with a diameter of 36\,$\mu$m. The modified sensor with the attached sample is glued to a cross made of Kevlar strings using Stycast 1299. With overall dimensions of 14\,×\,15\,×\,13.3\,mm$^3$, the TM-dHvA assembly fits into most conventional cryostats and commercially available rotation stages. All measurements presented here were performed using an Oxford Instruments HelioxVL $^{\mathrm{3}}$He cryostat with a cooling power of approximately 40\,$\mu$W at 300\,mK.


\begin{figure}[!tb]
\includegraphics{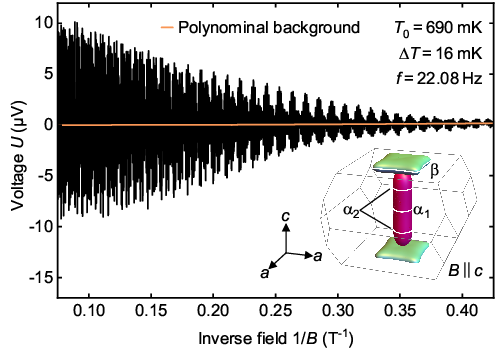}
\caption{Quantum oscillations, Fermi surface and extremal orbits of \MoSi\ for $\mathbf{B} \parallel \mathbf{c}$. The raw demodulated voltage $U$ at $T_{0}=690$\,mK and $f=22.08$\,Hz is shown as a function of $1/B$. The non-oscillatory background is negligible compared to the quantum oscillation amplitudes. The Fermi surface of \MoSi\ exhibits one extremal electron orbit $\beta$, one minimal hole orbit \Aone\ and two symmetry-equivalent maximal hole orbits \Atwo.\label{Fig:MoSi2}}
\end{figure}

\begin{figure*}[!tb]
\includegraphics[width=\textwidth]{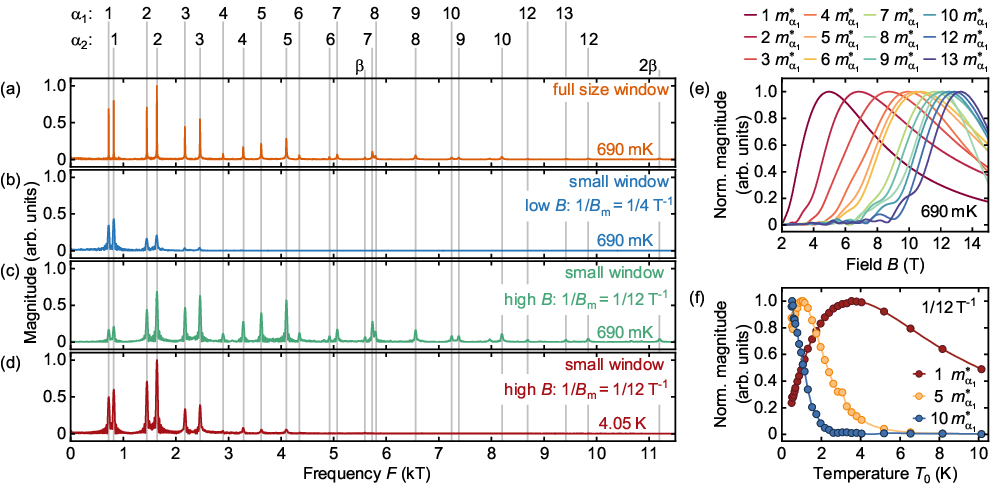}
\caption{Field and temperature evolution of TM-dHvA oscillation amplitudes in \MoSi. (a)~FFT spectrum of the data in Fig.\,\ref{Fig:MoSi2}(a) over a large window between 1/15 and 1/2\,T$^{-1}$. Frequencies \Aone, \Atwo\ and $\beta$ and harmonic content up to $p=13$ are indicated by vertical lines. The $p^{\mathrm{th}}$ harmonics of \Aone\ and \Atwo\ correspond to orbits with masses of $p\,m^*_{\alpha_{1,2}}$. (b)~FFT spectrum for a narrow window of 1/30\,T$^{-1}$ width centered at $1/B_{\mathrm{m}}=1/4$\,T$^{-1}$ and $T_0=690$\,mK. Here, signals from $1\alpha_{1,2}$ dominate the spectrum. (c) As in (b) but with the window centered at $1/B_{\mathrm{m}}=1/12$\,T$^{-1}$. Here, the orbits with large masses dominate, and the light fundamentals are suppressed. (d)~As in (c) but at $T_0=4.05$\,K. The frequencies corresponding to larger masses are again suppressed, and the fundamental peaks reemerge. (e)~Normalized IFFT amplitude at 690\,mK corresponding to \Aone\ and its harmonic content versus $B$. The amplitude peaks below 5\,T for $1\,m^*_{\alpha_1}$ and then decays to less than 20\% of the peak value at 15\,T. With increasing mass, amplitude onset and peak positions shift to higher $B$, with masses above $10\,m^*_{\alpha_1}$ only contributing appreciably at $B>10$\,T. (f)~Normalized IFFT amplitude for $1/B_{\mathrm{m}}=1/12$\,T$^{-1}$ corresponding to \Aone\ and its harmonic content versus \meanT. At temperatures above 4\,K the light mass $1\,m^*_{\alpha_1}$ dominates. At temperatures below 1\,K, $1\,m^*_{\alpha_1}$ is strongly suppressed, while $10\,m^*_{\alpha_1}$ approaches its peak value.\label{Fig:MoSi2_selectivity}}
\end{figure*}

\deleted{\MoSi\ crystallizes in the centrosymmetric tetragonal space group $I4/mmm$ (SG 139) with Mo on Wyckoff position 2a and Si on Wyckoff position 4e with $z=0.3353$. The band structure and Fermi surface were calculated with WIEN2k \cite{blahaWIEN2kAPW+loProgram2020} using the generalized gradient approximation of Perdew, Burke, and Ernzerhof \cite{perdewGeneralizedGradientApproximation1996} and including the effects of spin-orbit coupling. Experimental lattice constants $a=3.2064$\,\AA\ and $c=7.8478$\,\AA\ were used for the calculations.}

The calculated Fermi surface \added{of \MoSi} is shown in Fig.\,\ref{Fig:MoSi2}. It consists of a pill-shaped elongated hole pocket around the $\Gamma$-point (red) and a pillow-shaped electron pocket around the Z-point of the Brillouin zone (green). For $\mathbf{B} \parallel \mathbf{c}$, the Fermi surface exhibits four extremal orbits. There is one minimal orbit \Aone\ around the hole pocket and two maximal orbits \Atwo\ which have the same dHvA frequency due to symmetry. Only one extremal orbit $\beta$ resides on the electron pocket.

We used temperature modulation to measure quantum oscillations of the magnetization in \MoSi\ with the magnetic field applied along the $c$-axis. Figure\,\ref{Fig:MoSi2} shows the raw demodulated voltage $U$ as a function of inverse magnetic field at $T_0=690$\,mK for $\Delta T=16$\,mK and $f=22.08$\,Hz. The data are characterized by quantum oscillations dominated by a large-amplitude beating pattern arising from two close-by frequencies without any appreciable non-oscillatory background.

Figure\,\ref{Fig:MoSi2_selectivity}(a) shows the Fast Fourier transform (FFT) of the data presented in Fig.\,\ref{Fig:MoSi2} over a large reciprocal field interval between 1/15 and 1/2\,T$^{-1}$. The spectrum is dominated by two strong fundamental frequencies at 723 and 819\,T, which correspond to the extremal orbits \Aone\ and \Atwo, respectively, in good agreement with literature \cite{vanruitenbeekHaasvanAlphenEffect1987,matinExtremelyLargeMagnetoresistance2018,pavlosiukGiantMagnetoresistanceFermisurface2022}. The FFT spectrum is rich in harmonic content of the orbits \Aone\ and \Atwo, which could be identified up to $p=13$, with the exception of $p=11$ of both orbits. The large oscillation amplitude and strong harmonic content arise from the low curvature of the cross-sectional area of the pill along the $c$-axis combined with a sizeable Zeeman splitting. The less prominent frequency at 5.60\,kT generated by the electron orbit $\beta$ has not been observed previously for $\mathbf{B} \parallel \mathbf{c}$ \cite{vanruitenbeekHaasvanAlphenEffect1987,pavlosiukGiantMagnetoresistanceFermisurface2022}. However, our results are in agreement with measurements close to the $c$-axis \cite{vanruitenbeekHaasvanAlphenEffect1987,pavlosiukGiantMagnetoresistanceFermisurface2022}.


To demonstrate the non-monotonic field and temperature evolution of the amplitudes, we computed FFTs at different \meanT\ with moving windows with a constant width of $1/30$\,T$^{-1}$. To compare the spectral weights, we used rectangular FFT windows and normalized each spectrum to \DT.

Figure\,\ref{Fig:MoSi2_selectivity}(b) and (c) show FFTs performed at $T_0=690$\,mK and at inverse mean fields $1/B_{\mathrm{m}}=1/4$\,T$^{-1}$ and $1/B_{\mathrm{m}}=1/12$\,T$^{-1}$, respectively. For the low-field window centered at $1/B_{\mathrm{m}}=1/4$\,T$^{-1}$, the fundamental frequencies \Aone\ and \Atwo\ dominate the spectrum. For the high-field window centered at $1/B_{\mathrm{m}}=1/12$\,T$^{-1}$, their amplitudes are strongly suppressed, while peaks at higher harmonic frequencies are enhanced. Figure\,\ref{Fig:MoSi2_selectivity}(d) shows the FFT for $1/B_{\mathrm{m}}=1/12$\,T$^{-1}$,  and $T_0=4.05$\,K. Compared to Fig.\,\ref{Fig:MoSi2_selectivity}(c), where signal from the harmonic content associated with heavy \mstar\ dominates, the fundamental peaks are present at 4.05\,K, while the amplitudes of higher harmonics are suppressed. It follows that the choice of field window position and temperature range determines the range of orbit masses \mstar\ that contribute to the FFT spectrum of the TM-dHvA oscillations.

To analyze the oscillation amplitudes of individual harmonics, we first computed FFTs using the large window from 1/15 to 1/2\,T$^{-1}$, isolated individual FFT peaks using a rectangular window, and transformed the signal back via an inverse FFT (IFFT). In the following, we focus on \Aone\ and its harmonic content, \Atwo\ and $\beta$ behave analogously. 

Figure\,\ref{Fig:MoSi2_selectivity}(e) shows the normalized IFFT oscillation amplitudes of \Aone\ and all identified harmonics $p=2,\ldots,13$ (excluding $p=11$) at 690\,mK as a function of magnetic field. Deviations of the field dependence from the \dRdT\ term, as described in Fig.\,\ref{Fig:LK}(d), arise from the additional factors in $M_{0}$ that scale with $B$ and the harmonic index $p$, such as the Dingle factor \cite{shoenbergMagneticOscillationsMetals1984,dingleMagneticPropertiesMetals1952}. However, these contributions increase monotonically with increasing $B$, so that the non-monotonic field dependence of the TM-dHvA amplitudes versus $B$ can be ascribed to \dRdT.  

The amplitude of the oscillations associated with $1\,m^*_{\alpha_1}$ reaches its maximum at a field below 5\,T and decreases for higher fields. With increasing mass, onset and peak positions shift systematically to higher fields. By selecting an appropriate field range, one can thus enhance or suppress contributions from specific cyclotron mass ranges.

Selecting a specific mass range by using the sample temperature \meanT\ is even more straightforward, because \RT\ is the only term in the Lifshitz--Kosevich formalism that depends explicitly on temperature. 

The temperature dependence of the oscillation amplitudes associated with \Aone\ is shown in Fig.\,\ref{Fig:MoSi2_selectivity}(f) for selected masses of 1, 5 and 10\,$m_{\alpha_1}$. The temperature behavior of each orbit was analyzed by evaluating the amplitudes of IFFTs shown in Fig.\,\ref{Fig:MoSi2_selectivity}(e) at different $T_0$, but fixed inverse field of $1/12$\,T$^{-1}$. \added{Fitting the \dRdT\ to the temperature dependence shown in Fig.\,\ref{Fig:MoSi2_selectivity}(f) allowed us to extract the cyclotron masses, yielding $m_{\alpha_1} = (0.362 \pm 0.009)\,m_{\mathrm{e}}$ and $m_{\alpha_2} = (0.372 \pm 0.009)\,m_{\mathrm{e}}$ and $m_{\beta} = (1.32 \pm 0.05)\,m_{\mathrm{e}}$.} The amplitude of the 1\,$m_{\alpha_1}$ orbit is increasingly suppressed below 4\,K. Below 1\,K it decays to $\approx 20\%$ of its peak value, while 10\,$m_{\alpha_1}$ approaches its maximum. This behavior is particularly advantageous because it allows the use of low temperatures to isolate\added{ and enhance} weak oscillatory contributions from heavy orbits\replaced{ that may not be resolved as individual frequency peaks (see the supplementary materials) or may be below the detection limit in a conventional dHvA experiment.}{ that may otherwise be masked by strong oscillations from light orbits in the same frequency range.}

\deleted{Fitting the \dRdT\ to the temperature dependence shown in Fig.\,\ref{Fig:MoSi2_selectivity}(f) allowed us to extract the cyclotron masses, yielding $m_{\alpha_1} = (0.362 \pm 0.009)\,m_{\mathrm{e}}$ and $m_{\alpha_2} = (0.372 \pm 0.009)\,m_{\mathrm{e}}$ and $m_{\beta} = (1.32 \pm 0.05)\,m_{\mathrm{e}}$.}

\added{Conceptually, cyclotron mass discrimination should be possible in other quantum oscillatory quantities scaling with \dRdT\ \cite{huangNonlinearShubnikovdeHaas2021,palaciomoralesThermoelectricPowerQuantum2016,mercureQuantumOscillationsMetamagnetic2010,xuCrystalGrowthQuantum2019} including the recently demonstrated $3\omega$ Shubnikov-de Haas technique, which relies on self-heating by the transport current \cite{huangNonlinearShubnikovdeHaas2021}. While mass discrimination was not explored in Ref.~\cite{huangNonlinearShubnikovdeHaas2021}, its implementation would presumably hinge on the accurate experimental measurement of the temperature modulation amplitude, as opposed to modeling the $T$-dependent prefactor of \dRdT. Apart from these technical issues, the measured transport versus thermodynamic quantities are complementary, with Shubnikov-de Haas oscillations tending to be dominated by small Fermi surfaces \cite{huangNonlinearShubnikovdeHaas2021}, whereas all Fermi surface sheets contribute to dHvA spectra (see the supplementary material for details).}


In conclusion, we developed a TM-dHvA measurement technique and detected quantum oscillations of \dMdT\ in the compensated semimetal \MoSi. Using a series of harmonics with masses in the range $\approx 0.3\,m_{\mathrm{e}}-4\,m_{\mathrm{e}}$  originating from the $\alpha$ orbits in \MoSi, we demonstrated that TM-dHvA allows for the isolation of orbits with heavy masses from orbits with light masses, which is not possible in conventional dHvA experiments.\deleted{ Importantly, this is feasible even when the corresponding orbits have the same frequency.} These properties are expected to be beneficial for quantum oscillation experiments in a wide range of material classes where heavy and light quasiparticles at the Fermi surface coexist. \\

\section*{Supplementary Material}

\added{See the supplementary material for crystal growth parameters, density functional theory calculations, numerical dHvA simulations showcasing mass discrimination and signal enhancement, and a comparison to the $3\omega$ transport technique.}

\section*{Conflict of Interest}
The authors have no conflicts to disclose.\\

\section*{Author Contributions}
\textbf{Michelle Hollricher:} Writing -- original draft (lead); Writing -- review \& editing (equal); Investigation (lead); Methodology (lead). \textbf{Andreas Bauer:} Resources (equal); Writing -- review \& editing (supporting). \textbf{Leo Maximov:} Resources (equal). \textbf{Louw Feenstra:} Data curation (supporting). \textbf{Christian Pfleiderer:} Conceptualization (equal); Supervision (supporting). \textbf{Marc A. Wilde:} Writing -- review \& editing (equal); Conceptualization (equal); Supervision (lead).

\section*{Acknowledgments}
We wish to thank T. Hesjedal for discussions when preparing this manuscript. We gratefully acknowledge financial support of the DFG via TRR360 (ConQuMat, Project No.\ 492547816), SPP2137 (Skyrmionics, Project No.\ 403191981, Grant PF393/19), DFG-GACR project WI3320/3 (Project No. 323760292), MCQST under Germany's Excellence Strategy EXC-2111 (Project No.\ 390814868) and the ERC through Advanced Grant No.\ 788031 (ExQuiSid).

\section*{Data Availability Statement}
The data that support the findings of this study are available from the corresponding author upon reasonable request.

\bibliography{2025_TMT_MoSi2}

\end{document}


\newcommand{\dMdT}{$\partial M/\partial T$}
\newcommand{\MoSi}{MoSi$_{2}$}
\newcommand{\RT}{$ R_{\mathrm{T}}$}
\newcommand{\dRdT}{$\partial R_{\mathrm{T}}/\partial T$}
\newcommand{\DT}{$ \Delta T $}
\newcommand{\meanT}{$ T_{0}$}
\newcommand{\mstar}{$m^{\ast}$}
\newcommand{\Aone}{$\alpha_{1}$}
\newcommand{\Atwo}{$\alpha_{2}$}
\newcommand{\fDT}{$ \Delta T \cdot f$}
\newcommand{\Rox}{RuO$_{2}$}


\title{Supplementary material: Cyclotron mass-selective de~Haas--van~Alphen measurements using temperature modulation}

\author{Michelle Hollricher}
\email{michelle.hollricher@tum.de}
\affiliation{School of Natural Sciences, Technical University of Munich, Garching, Germany}
\affiliation{Zentrum f\"ur QuantumEngineering (ZQE), Technical University of Munich, Garching, Germany}

\author{Andreas Bauer}
\affiliation{School of Natural Sciences, Technical University of Munich, Garching, Germany}
\affiliation{Zentrum f\"ur QuantumEngineering (ZQE), Technical University of Munich, Garching, Germany}

\author{Leo Maximov}
\affiliation{School of Natural Sciences, Technical University of Munich, Garching, Germany}
\affiliation{Zentrum f\"ur QuantumEngineering (ZQE), Technical University of Munich, Garching, Germany}

\author{Louw Feenstra}
\affiliation{School of Natural Sciences, Technical University of Munich, Garching, Germany}
\affiliation{Fakult\"at für Physik, Ludwig-Maximilians-Universit\"at, Munich, Germany}

\author{Christian Pfleiderer}
\affiliation{School of Natural Sciences, Technical University of Munich, Garching, Germany}
\affiliation{Zentrum f\"ur QuantumEngineering (ZQE), Technical University of Munich, Garching, Germany}
\affiliation{Heinz Maier-Leibnitz Zentrum (MLZ), Technical University of Munich, Garching, Germany}
\affiliation{Munich Center for Quantum Science and Technology (MCQST), Technical University of Munich, Munich, Germany}

\author{Marc A. Wilde}
\email{marc.wilde@tum.de}
\affiliation{School of Natural Sciences, Technical University of Munich, Garching, Germany}
\affiliation{Zentrum f\"ur QuantumEngineering (ZQE), Technical University of Munich, Garching, Germany}

\date{\today}

\begin{abstract}
In this supplementary material, we first describe briefly the crystal growth and density functional theory calculations of \MoSi\ and then provide numerical simulations within the Lifshitz-Kosevich formalism to illustrate the possible benefits of temperature-modulated dHvA experiments. For systems that exhibit overlapping or identical dHvA frequencies with strongly different effective masses, we showcase specific minimal examples of two orbits differing in mass by two orders of magnitude and  demonstrate how conventional Fast Fourier Transform analysis may fail to determine -- or even detect -- the oscillatory component with heavy mass when the frequency difference is smaller than the width of the FFT peaks. We then show that TM-dHvA can unambiguously disentangle the two contributions. We further comment on different dynamic range and signal-to-noise constraints in conventional and TM-dHvA experiments. Finally, we give a brief comparison to the $3\omega$-SdH technique.
\end{abstract}

\maketitle

\section{Crystal growth}

Starting from stoichiometric amounts of Mo (3N8) and Si (6N), we first prepared polycrystalline rods using an arc-melting furnace, followed by an inductively heated rod casting furnace \cite{bauerUltrahighVacuumCompatible2016}. Single-crystal growth was carried out in a high-temperature floating-zone furnace at a rate of 10 mm/h under an argon atmosphere of 10 bar \cite{bauerCompositionalStudiesMetals2022}. The crystal was oriented using Laue X-ray diffraction and subsequently cut with a wire saw. The sample used in the experiment is a 1\,mm cube of single-crystal MoSi$_2$ with faces oriented along the c, a, and b axes.

\section{Density functional theory calculations}

\MoSi\ crystallizes in the centrosymmetric tetragonal space group $I4/mmm$ (SG 139) with Mo on Wyckoff positions 2a and Si on Wyckoff positions 4e with $z=0.3353$. The band structure and Fermi surface were calculated with WIEN2k \cite{blahaWIEN2kAPW+loProgram2020} using the generalized gradient approximation of Perdew, Burke, and Ernzerhof \cite{perdewGeneralizedGradientApproximation1996} and including the effects of spin-orbit coupling. Experimental lattice constants $a=3.2064$\,\AA\ and $c=7.8478$\,\AA\ were used for the calculations.

\section{Numerical simulations of overlapping dHvA frequencies with strongly different quasiparticle masses}

In the following, we assume negligible Dingle damping and spin damping for simplicity. Note that the inclusion of additional damping factors is straightforward. We consider minimal examples of two dHvA orbits, one with a mass of $10\,m_\mathrm{e}$ and one with a mass of $0.1\,m_\mathrm{e}$, and systematically vary the frequencies $F_i$ and relative phases $\Delta \phi=\phi_i-\phi_{i-1}$. We simulate experimental magnetization data using $M(1/B)=\sum_{i}M_i$ with the $M_i$ calculated within the Lifshitz-Kosevich formalism
\begin{equation}
    M_i=-\left(\frac{e}{\pi }\right)^{5/2} \frac{F_i}{m_i^*}  \sqrt{\frac{B}{2 \hbar  A_i^{\prime \prime}}} \sum_{p=1}^{\infty} p^{-3/2} \frac{ 2 \pi ^2 p m_i^*  k_B T }{\hbar e B \sinh{\frac{2 \pi ^2 p m_i^*  k_B T}{\hbar e B}}}\sin \left(2 \pi p \left(\frac{F_i}{B}-\frac{1}{2}\right)+\phi_i \right)  \mbox{ ,}
    \label{eq:LK-formalism}
\end{equation}
with $A^{\prime \prime}$ being the curvature of the cross-sectional area along the magnetic field direction. We incorporate the \emph{ad hoc} phases $\phi_i$ to describe the net result of various experimental phase contributions, including $\mp\frac{\pi}{4}$ for a maximal/minimal orbit, Berry phases, and spin damping. Hann windows from $B_{\mathrm{min}}=10$\,T to $B_{\mathrm{max}}=15$\,T and zero padding are applied to the $M(1/B)$ data, followed by fast Fourier transforms, as might be done in a real experiment. We note, however, that the main results obtained in this supplementary material are independent of the details of the data treatment. This reciprocal field window size of $1/30$\,T$^{-1}$ corresponds to a frequency $F_w=30$\,T which sets the scale for the minimum difference of two frequencies that is resolvable in experiment as $\Delta F \approx c F_w$, where $c$ is a factor of order unity that depends on the windowing function. Intuitively, this may be understood as the requirement that there be more than one full cycle of the difference frequency within the reciprocal field window in order to accurately sample the beating pattern. In our computer experiment, we thus expect to resolve two individual frequencies for $\Delta F \gtrsim 100$\,T.

\subsection{Overlapping frequencies with strongly different masses in conventional dHvA experiments}

In this section, we show that in conventional dHvA experiments it is usually not possible to determine the existence and cyclotron mass of a heavy-mass orbit, when its FFT peak overlaps with the FFT peak of one or more light-mass orbits such that the individual maxima are not resolved in the spectrum.

\begin{figure}[!tb]
\centering
\includegraphics{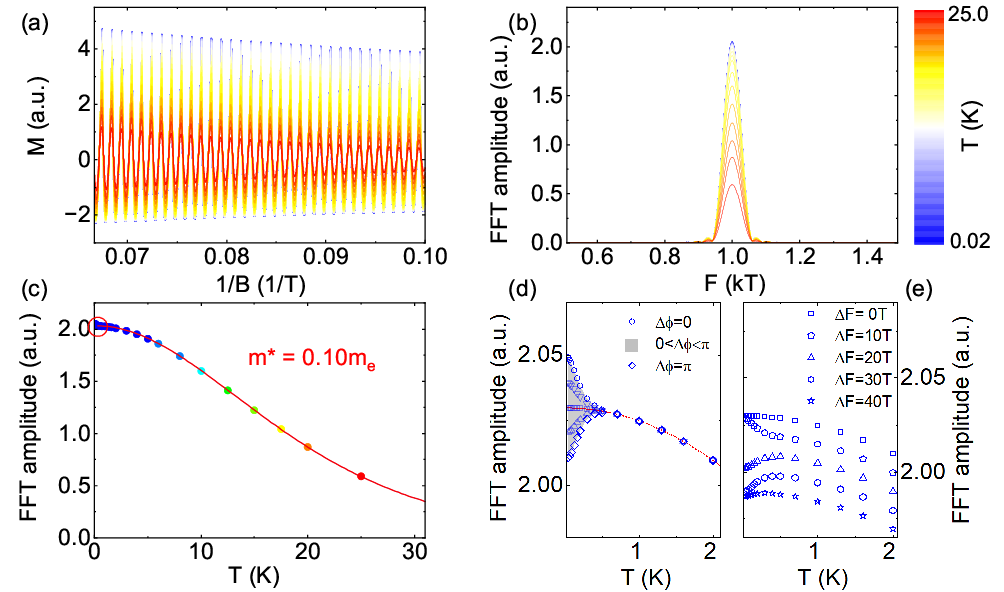}
\caption{Simulated conventional dHvA experiment with two overlapping frequencies around $F=1$\,kT of varying phases with masses of $0.1\,m_\mathrm{e}$ and $10\,m_\mathrm{e}$, respectively (a) DHvA amplitude vs $1/B$ for temperatures in the range of $T=0.02-25$\,K, for $F_1=F_2=1$\,kT and $\Delta \phi=0$ (b)~FFT of the data in (a) with a Hann window between $1/15-1/10$\,$T^{-1}$ (c)~FFT amplitude vs $T$ and conventional single-component LK-fit (red) yielding the mass $m^*=0.10\,m_\mathrm{e}$ of the light orbit only. The slight upturn at the lowest $T$ (red circle) is likely to be overlooked in a real experiment. (d) FFT amplitude vs $T$ in the low temperature regime for different relative phases between $\Delta \phi=0$, and $\Delta \phi=\pi$ (as, e.g., for one orbit with a $\pi$ Berry phase). Depending on the relative phase, data points can follow LK-curves in the shaded region (e) FFT amplitude for a minimal and a maximal orbit differing slightly in frequency by $\Delta F=10,20,30,40$\,T, again resulting in different line shapes that do not follow a simple behavior. \label{Fig:SupplementaryFig1}
\vspace{0.25cm} \\
ALT TEXT: Simulated conventional de Haas-van Alphen experiment for two orbits at the same frequency with different cyclotron masses, demonstrating that the heavy-mass component is largely undetectable by standard Lifshitz-Kosevich fitting.}
\end{figure}

In general, the individual frequency contributions will differ in both frequency and phase. For the sake of clarity, let us first consider the somewhat fine-tuned but straightforward case of two extremal cyclotron orbits with the same phase and same frequency $F_1=F_2=F=1$\,kT, one with a light mass of $0.1\,m_\mathrm{e}$ and one with a heavy mass of $10\,m_\mathrm{e}$ being present in the same system. Subsequently, we will independently vary the relative phase and frequency differences and discuss the experimental signatures of these variations. Figure~\ref{Fig:SupplementaryFig1}(a) shows the amplitude evolution of the resulting dHvA oscillations in the reciprocal magnetic field window for different temperatures in the range of $0.02-25$\,K (color scale). The two individual contributions cannot be discerned in the raw data. The same is true for the FFT peaks depicted in Fig.~\ref{Fig:SupplementaryFig1}(b) for different temperatures, which show no obvious signs of their dual nature. For the effective mass analysis, the FFT amplitude is usually plotted versus temperature and fitted with the Lifshitz-Kosevich (LK) factor as shown in Fig.~\ref{Fig:SupplementaryFig1}(c). Here, a minute upturn in the curve at the lowest temperatures (red circle) may be discerned; however, in practice and in the presence of experimental uncertainties, this would likely be overlooked, and a single-component LK-fit (red line) yields the mass of the light orbit $m^*=0.1\,m_\mathrm{e}$ only.

If the upturn is identified as an additional heavy-mass component by the able experimenter, a two-component LK-fit may be attempted. However, the amplitudes of the two oscillations at the same frequency are generally not additive when a relative phase $\Delta \phi$ between them is taken into account, as we demonstrate in Fig.~\ref{Fig:SupplementaryFig1}(d,e): when varying the phase difference, the amplitude evolution covers the gray shaded area in Fig.~\ref{Fig:SupplementaryFig1}(d) that is bounded by the in-phase case $\Delta \phi=0$ (circles), where the amplitudes add, and the anti-phase case $\Delta \phi=\pi$ (diamonds), where the amplitudes subtract. Note that this includes the case of no amplitude change at all due to the heavy-mass component being phase-shifted by $\Delta \phi=\pi/2$. Because the relative phase is, in general, not known in the experiment and exceedingly hard to determine in the case of oscillations far from the quantum limit with vastly different amplitudes, the relative phase must be incorporated as a fit parameter in addition to the amplitudes and masses, yielding ambiguous results especially for relative phases close to $\pm \pi/2$. We conclude that analysis of the heavy-mass component is not generally possible even in this simplified scenario with exactly identical frequencies.  

Figure~\ref{Fig:SupplementaryFig1}(e) shows the situation for differing frequencies corresponding to a maximal and a minimal orbit, i.e., for $\Delta \phi=\pi/2$. If that difference in frequency $\Delta F$ is on the order of the field window $F_w=30$\,T calculated above, it will show up as an effective smearing of the relative phase, without resolving into two individual frequency peaks. This field-dependent smearing results in a non-monotonic amplitude behavior vs temperature depending on the value of $\Delta F$. In addition, in this case of overlapping non-identical frequency peaks, the result depends sensitively on the experimental field window.

In a real experiment, quantum oscillations with overlapping FFT peaks will, in general, have both different frequencies and different phases. Extracting this information from the temperature dependence of a single FFT maximum is not feasible in practice without additional information.

\subsection{Overlapping frequencies with strongly different masses in temperature-modulation dHvA experiments }

We now consider the signatures of the same simulated dHvA orbits as above as they would be measured in temperature-modulation dHvA (TM-dHvA) experiments. The quantum oscillations in the TM-dHvA signal are proportional to \dMdT $=\sum_i \partial M_i / \partial T$ with
\begin{equation}
    \frac{\partial M_i}{\partial T} \sim \frac{F_i}{m_i^*} \, \frac{\partial R_{\mathrm{T,i}}}{\partial T} \, \sin \left[ 2\pi \left( \frac{F_i}{B} - \frac{1}{2} \right) +  \phi_i \right] \mbox{ ,}
\end{equation}
where we restricted the discussion to the fundamental frequency for the sake of clarity. Inclusion of higher harmonic terms is straightforward and does not alter the line of argument.

\begin{figure}[!tb]
\centering
\includegraphics{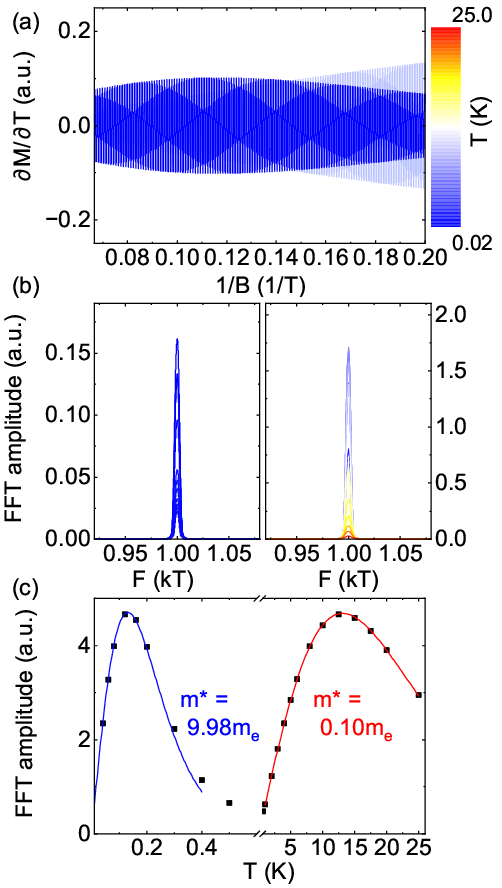}
\caption{Simulated TM-dHvA experiment demonstrating the separation of a heavy-mass and a light-mass orbit into two independent temperature regimes. The data can be evaluated separately, irrespective of the exact frequencies and relative phases. (a)~TM-dHvA signal of the sum of a $10\,m_\mathrm{e}$ and a $0.1\,m_\mathrm{e}$ orbit at $T=0.02$\,K (dark blue) and $T=4$\,K (light blue). Only the fundamental terms with $p=1$ are shown for clarity. (b) Fourier transforms for $T<1$\,K (left panel) and for $T>1$\,K (right panel). (c) FFT amplitude vs. $T$. In the low-temperature regime, the amplitude is dominated by the large-mass component. A fit of $\partial R_\mathrm{T} / \partial T$ (blue line) to the data below $0.4$\,K yields the correct effective mass of the heavy component close to $10\,m_\mathrm{e}$. A fit to the high-temperature data above $1$\,K (red line) yields $m^*=0.1\,m_\mathrm{e}$, the mass of the light component.   \label{Fig:SupplementaryFig2}
\vspace{0.25cm} \\
ALT TEXT: Simulated temperature-modulated de Haas-van Alphen experiment for the same heavy and light mass orbit, which separate into distinct low- and high-temperature regimes and can each be evaluated independently.}
\end{figure}

Figure~\ref{Fig:SupplementaryFig2}(a) shows the fundamental Fourier component of the quantum oscillations in \dMdT\ for $F_1=F_2=1$\,kT at two different temperatures $T=0.02$\,K (dark blue) and $T=4$\,K (light blue). Due to the mass-selective properties, the low-temperature quantum oscillations primarily arise from the heavy-mass component, whereas the high-temperature oscillations arise from the light-mass component. Note that the oscillation envelopes peak at different field values, allowing for a selective addressing via the choice of field window. However, to ease comparison with the conventional dHvA data discussed above, we focus on mass selectivity on the temperature axis here, and analyze the data at the same field value $B_{m}=2 B_{min}B_{max}/(B_{min}+B_{max})=12$\,T used above. The Fast Fourier transforms for $T<1$\,K and for $T>1$\,K are shown in the left and the right panel of Fig.~\ref{Fig:SupplementaryFig2}(b), respectively. The striking difference between the two temperature regimes becomes even more apparent when plotting the FFT peak amplitude versus temperature in Fig.~\ref{Fig:SupplementaryFig2}(c): Two peaks, each individually following a separate \dRdT\ behavior on the temperature axis, are observed. The two peaks in the two regimes can be fitted individually with \dRdT\ curves (blue and red lines), and the masses can be extracted separately. Here, the slight deviation of the fitted heavy mass value of $0.2\%$ is due to the residual contribution of the light-mass orbit.

This TM-dHvA analysis is fully independent of relative phase and small frequency differences, as discussed for the conventional dHvA case, because the two signals are separated on the temperature axis by their different cyclotron masses.

In summary, when FFT peaks arising from cyclotron orbits with strongly different masses are not resolved individually in frequency space in a conventional dHvA experiment, the analysis of the heavy component is usually not possible and might even evade detection altogether. In TM-dHvA, however, the two contributions can be separated on the temperature (and magnetic field) axis, with the process becoming increasingly efficient as the mass difference increases.

\section{Remarks on dynamic range and signal-to-noise limitations of conventional dHvA and TM-dHvA experiments}

The enhancement of heavy-mass quantum oscillations at low temperatures, accompanied by the simultaneous suppression of light-mass oscillations, is a generic feature of the TM-dHvA effect and thus also relevant in cases where the frequency peaks are well separated. While in this case, the heavy-mass orbits are not "masked" by the light ones, they might still evade detection due to the finite dynamic range, or simply the noise floor, of the detection system.

The quantum oscillation amplitudes of two orbits are expected to differ by a factor $m^*_1/m^*_2$ for similar frequency values (see Eq.\,\ref{eq:LK-formalism}) in the absence of Dingle damping. Since Dingle damping due to lifetime broadening of the Landau levels scales exponentially with the ratio of the lifetime broadening to the cyclotron energy (inversely proportional to the quasiparticle mass), heavy-orbit oscillations tend to occur at much higher fields only, with their amplitude strongly reduced beyond the mass ratio. 

In such situations, the strong enhancement of the heavy-mass amplitude in TM-dHvA may render oscillations observable, whose amplitude would otherwise be below (i) the absolute noise floor of a given detection system or (ii) below the minimal signal amplitude dictated by the dynamic range of the detection system in the presence of the high-amplitude quantum oscillations of the light orbit. The maximum enhancement of the \dRdT\ factor to values larger than one increases towards lower temperatures without bounds. In practice, the achievable enhancement will be limited by the lowest experimental temperature and by the Dingle factor.

A direct quantitative comparison between TM-dHvA and dHvA in simulations, however, would not be meaningful because the setup-dependent achievable noise floor and dynamic range of a conventional dHvA and a TM-dHvA setup would need to be defined on an \emph{ad hoc} basis.

\section{Remarks on TM-dHvA as compared to $\bm{3\omega}$-SdH}

The cyclotron-mass selectivity applies, in principle, to all quantum oscillations that scale with \dRdT, including the recently demonstrated $3\omega$-Shubnikov-de Haas (3$\omega$-SdH) technique, relying on self-heating by the transport current \cite{huangNonlinearShubnikovdeHaas2021}. While mass discrimination was not explored in Ref.~\cite{huangNonlinearShubnikovdeHaas2021}, its implementation would presumably hinge on the accurate experimental measurement of the temperature modulation amplitude, as opposed to using the Wiedemann-Franz law to approximately account for the $T$-dependence of the sample's thermal conductivity and staying away from the regime where the $T$-dependence of the specific heat becomes important.

On a technical level, the $3\omega$-SdH technique needs sophisticated microstructures fabricated by focused-ion-beam (FIB) milling, which is not necessary for TM-dHvA, which works on plain bulk samples. However, both variants may be advantageous or disadvantageous: FIB milling can be detrimental to the sample properties, whereas a lack of large single crystals would favor microstructures.

Apart from these technical issues, the measurement of transport quantities is complementary to the measurement of thermodynamic quantities such as the magnetization: While the dHvA oscillations occur in a thermodynamic quantity directly related to the thermodynamic potentials, and thus to the density of states, the SdH effect occurs in the resistivity of a sample, and can thus only be understood quantitatively, when the microscopic details of scattering in a non-equilibrium situation are taken into account. As a consequence, SdH favors the charge carriers that dominate conduction, while dHvA samples all $k$-space contributions to the density of states at the Fermi level on an equal footing. SdH oscillations thus tend to be sensitive to small orbits only, due to the limited phase space for scattering.

Another point of practical relevance is the potential suppression of non-oscillatory backgrounds of the measured quantities. While this may seem like an unimportant detail, large backgrounds with complex shapes can impose constraints on signal analysis, especially at low frequencies, where imperfect background subtraction can lead to spurious signals. In addition, the dynamic range of the measurement apparatus may also play a role here. While $3\omega$-SdH was shown to suppress large non-oscillatory magnetoresistance contributions, occurring especially in compensated semimetals, TM-dHvA may suppress large non-oscillatory magnetic backgrounds in magnetic systems. In both cases, this depends on the backgrounds having a weaker temperature dependence than the quantum oscillations, so the advantages are expected to be highly material-dependent.

\bibliography{2025_TMT_MoSi2}